\documentclass{mn2e} 
\usepackage{graphics}
\usepackage{epsfig}

\def\etal{et\thinspace al.\ }                               
 
\def\beq{\begin{equation}}                          
\def\eeq{\end{equation}}                              
\def\beqa{\begin{eqnarray}}                         
\def\eeqa{\end{eqnarray}}                             
\def\beqan{\begin{eqnarray*}}                      
\def\eeqan{\end{eqnarray*}}                          

\newbox\grsign \setbox\grsign=\hbox{$>$} \newdimen\grdimen \grdimen=\ht\grsign
\newbox\simlessbox \newbox\simgreatbox
\setbox\simgreatbox=\hbox{\raise.5ex\hbox{$>$}\llap
     {\lower.5ex\hbox{$\sim$}}}\ht1=\grdimen\dp1=0pt
\setbox\simlessbox=\hbox{\raise.5ex\hbox{$<$}\llap
     {\lower.5ex\hbox{$\sim$}}}\ht2=\grdimen\dp2=0pt

\title[The CaT strength in Seyfert nuclei revisited]{The CaT strength
  in Seyfert nuclei revisited: analyzing young stars and non-stellar
  light contributions to the spectra}

\author[Vega \etal]
       {L. R. Vega$^{1,2}$\thanks{E-mail: luisv@mail.oac.uncor.edu},
	 N. V. Asari$^{3}$\thanks{E-mail: natalia@astro.ufsc.br}, 
	 R. Cid Fernandes$^{3}$\thanks{E-mail: cid@astro.ufsc.br},
	 A. Garcia-Rissmann$^{4}$\thanks{E-mail: agarciar@eso.org},
	 \newauthor
	 T. Storchi-Bergmann$^{5}$\thanks{thaisa@if.ufrgs,br},
	 R. M. Gonz\'alez Delgado$^{6}$\thanks{E-mail: rosa@iaa.es},
	 H. Schmitt$^{7}$\thanks{E-mail: henrique.schmitt@ nrl.navy.mil}\\
	 $^{1}$ Instituto de Astronom\'{\i}a Te\'orica y Experimental,
         IATE - Observatorio Astron\'omico, Universidad Nacional de
         C\'ordoba\\ Laprida 854, X5000BGR C\'ordoba, Argentina\\
	 $^{2}$ Secyt, Secretar\'{i}a de Ciencia y T\'ecnica de la
         Universidad Nacional de C\'ordoba, C\'ordoba, Argentina\\
	 $^{3}$ Universidade Federal de Santa Catarina,
         Florian\'opolis, Brazil\\
	 $^{4}$ European Southern Observatory, Karl-Schwarzschild Str. 2, D-85748 Garching b. M\"unchen, Germany\\
	 $^{5}$ Instituto de F\'{\i}sica, Universidade Federal do Rio
         Grande do Sul, C.P. 15001, 91501-970, Porto Alegre, RS,
         Brazil\\
         $^{6}$ Instituto de Astrof\'{\i}sica de Andaluc\'{\i}a
       (CSIC), P.O. Box 3004, 18080 Granada, Spain\\
         $^{7}$ Remote Sensing Division, Naval Research Laboratory,
       Washington, DC 20375; and Interferometrics, Inc., Herdon, VA 20171, USA}

\begin{document}

\maketitle

\begin{abstract} 
In a former paper (Garcia-Rissmann \etal 2005; hereafter Paper I), we
have presented spectra of 64 active, 9 normal and 5 Starburst galaxies
in the region around the near-IR Calcium triplet absorption lines
(CaT) and the [SIII]$\lambda$9069 line. In the present paper we
analyze the CaT strength ($W_{\rm CaT}$), and kinematical products
derived in that study, namely stellar ($\sigma_{\star}$) and ionized
gas ($\sigma_{\rm gas}$) velocity dispersions. Our main results may be
summarized as follows: (1) Seyfert 2s show no sign of dilution in
$W_{\rm CaT}$ with respect to the values spanned by normal galaxies,
even when optical absorption lines such as the CaII K band at 3933
\AA\ are much weaker than in old, bulge-like stellar populations. (2)
The location of Seyfert 2s in the $W_{\rm CaT}$-$W_{\rm CaK}$ plane is
consistent with evolutionary synthesis models. The implication is that
the source responsible for the dilution of optical lines in these AGN
is a young stellar population, rather than an AGN featureless
continuum, confirming the conclusion of the pioneer study of
Terlevich, D\'{\i}az \& Terlevich. (3) In Seyfert 1s, both $W_{\rm
  [SIII]}$ and $W_{\rm CaT}$ tend to be diluted due to the presence of
a non-stellar component, in agreement with the unification
paradigm. (4) A comparison of $\sigma_\star$ with $\sigma_{\rm gas}$
(obtained from the {\it core} of the [SIII] emitting line) confirms
the existence of a correlation between the typical velocities of stars
and clouds of the Narrow Line Region. The strength and scatter around
this correlation are similar to those previously obtained from the
[OIII]$\lambda$5007 line width.

\end{abstract}

\section{Introduction}

\label{sec:Introduction}

The stellar absorption lines from the near-IR Calcium triplet (CaT)
made their debut in the field of active galactic nuclei (AGN) with the
work of Terlevich, D\'{\i}az \& Terlevich (1990, hereinafter TDT).  At
that time, it was thought that the optical spectrum of Seyfert 2s
contained a featureless continuum (FC) from the AGN, which, in light
of unification scenarios (Antonucci 1993), should be associated with
scattered light from the hidden Seyfert 1 nucleus (Cid Fernandes \&
Terlevich 1995).  TDT realized that this idea could not be fully
correct, since their data indicated that, unlike for absorption lines
in the optical, the strength of the CaT in these sources shows no
signs of dilution with respect to the values found in normal
galaxies. Their interpretation of this fact was that what was called a
``non-stellar FC'' was in fact a stellar component associated to young
stellar populations. This conclusion was thoroughly confirmed by
numerous studies in the past decade (see Cid Fernandes \etal 2004 for
a review).

Since then, interest in the CaT shifted towards its use as a tracer of
stellar kinematics, particularly in AGN.  The location of the CaT in a
relatively clean spectral region makes it an ideal feature to measure
stellar velocity dispersions ($\sigma_\star$), which trace the
gravitational potential of the host galaxy's bulge.  In an influential
work, Nelson \& Whittle (1996; NW) used the CaT in a comparison
between stellar and gaseous kinematics of AGN, finding that the
typical velocity Narrow Line Region (NLR) clouds ($\sigma_{\rm gas}$)
correlates with $\sigma_\star$, albeit with significant scatter.  More
recently, a strong correlation between $\sigma_\star$ and black-hole
mass (Tremaine \etal 2002 and references therein) indirectly boosted
interest in observations of the CaT (Nelson 2000; Botte \etal 2004).

For completely different reasons, recent work on normal galaxies has
also raised the interest in the CaT. Empirical investigations have
revealed rather puzzling behavior of the CaT in bulges and elliptical
galaxies. Firstly, detailed population synthesis models tend to
over-predict the CaT strength (Saglia \etal 2002; Cenarro \etal 2003,
2004), particularly for giant ellipticals, though for dwarf
ellipticals the match between data and models is satisfactory
(Michielsen \etal 2007).  Secondly, while classic metallicity tracers
like the Mg$_2$ index are known to correlate with $\sigma_\star$ (eg,
Terlevich \etal 1981), tracing the well-known mass-metallicity
relation, the strength of the CaT appears to be {\it anti-correlated}
with $\sigma_\star$ (Cenarro \etal 2003, 2004; Falc\'on-Barroso \etal
2003; Michielsen \etal 2003). The interpretation of these results is
still not clear, and the situation is likely to be even more complex
in systems with varied star formation histories like AGN (Cid
Fernandes \etal 2004; Wild \etal 2007).

One thus sees that, more than 15 years after TDT introduced the CaT in
AGN research, there are still plenty of reasons to keep studying
it. With this general motivation, in Paper I we have carried out a
spectroscopic survey of Seyfert galaxies in the region including the
CaT and the [SIII]$\lambda$9069 emission line. Paper I concentrated in
the presentation of the data and the derivation of four main data
products: stellar velocity dispersions ($\sigma_\star$), [SIII]
emission line widths ($\sigma_{\rm [SIII]}$, which is representative
of the highly ionized component of the gas of the Narrow Line Region
in AGNs), [SIII] equivalent widths ($W_{\rm [SIII]}$), and the CaT
strength ($W_{\rm CaT}$).  Here we use these data to address the
following issues: (1) examine the contribution of a non-stellar
component in the near-IR spectra of Seyfert galaxies and test the
consistency of $W_{\rm [SIII]}$ and $W_{\rm CaT}$ data with the
unified model, (2) study the sensitivity of $W_{\rm CaT}$ to stellar
population properties, (3) investigate the connection between NLR and
stellar kinematics and CaT strength, and (4) report results of
spatially resolved spectroscopy, not reported in Paper I.

To study these questions, we complement the data in Paper I with
values of the equivalent width of the CaII K line ($W_{\rm CaK}$)
obtained from previous studies. As shown by Cid Fernandes \etal
(2001), the CaII K line is a powerful tracer of stellar populations
even in type 2 Seyferts. By comparing $W_{\rm CaT}$ with $W_{\rm CaK}$
we can assess whether the CaT is a useful stellar population tracer,
and determine whether the combination of CaK and CaT strengths is
explained as due to stars alone or if an FC component is necessary. We
also add in literature information on the width ($\sigma_{\rm
  [OIII]}$) of the [OIII]$\lambda$5007 line, which is useful to test
whether our results are somehow affected by the choice of
[SIII]$\lambda$9069 as a tracer of NLR motions.

This paper is organized as follows. Section
\ref{sec:CaT_NonStellarLight} presents a series of studies related to
the CaT strength. After showing CaT observational properties in
\ref{subsec:CaT in Sy}, we compare $W_{\rm CaT}$ with $W_{\rm CaK}$ in
\ref{subsec:Dilution} and investigate whether the location of the
points in the $W_{\rm CaT}$ versus $W_{\rm CaK}$ diagram is consistent
with the existence of an FC at near-IR wavelengths. In \ref{Stellar
  Models} we use evolutionary synthesis models to track the behavior
of both $W_{\rm CaT}$ and $W_{\rm CaK}$ as a function of age and
metallicity, and overlay the models onto the data in the $W_{\rm CaT}$
versus $W_{\rm CaK}$ diagram. In \ref{subsec:Spatial CaT Profiles} we
study the spatial behavior of $\sigma_{\star}$ and $W_{\rm CaT}$ for
about half of our sample. In Section \ref{sec:Kinematics} we analyze
stellar and ionized gas kinematics: stellar velocity dispersions and
its relation to CaT are studied in \ref{subsec:Stellar velocity
  dispersions}, while in \ref{subsec:Stellar vs. NLR Kinematics} we
discuss the link between stellar and gas kinematics. Finally, section
\ref{sec:Summary} summarizes our main results.

\section{The CaT strength: Effects of non-stellar light}

\label{sec:CaT_NonStellarLight}

One of our goals is to evaluate the usefulness of the CaT as a stellar
population diagnostic. Since our sample is dominated by AGN, before
discussing stellar populations we must first examine to which extent
non-stellar light affects our CaT measurements.

As explained in Paper I, we adopt the $W_{\rm CaT}$ and $W_{\rm CaT*}$
definitions of Cenarro et al (2001a). They offer two definitions of
the CaT equivalent width: `CaT' (which we call $W_{\rm CaT}$), which
consists of a sum of the equivalent widths of all three CaT lines, and
`CaT*' (called $W_{\rm CaT*}$ here), which corrects $W_{\rm CaT}$ for
contamination by Paschen line absorption.  These equivalent widths are
measured with respect to a continuum defined by fitting the spectrum
in 5 windows in the 8474-8792 \AA\ range. Because of the presence of
unwanted features in the observed spectra, we opted to measure $W_{\rm
  CaT}$ and $W_{\rm CaT*}$ on the synthetic spectra. On the other
hand, our spectral base used to fit the observed spectra does not have
stars with the spectral types with Paschen lines, so we use $W_{\rm
  CaT}$ as equivalent widths measurements for our spectra.

All nuclear CaT data refers to appertures of $\sim$ 2'' $\times$ 2''
(Paper I). To take into account the different velocity dispersions of
our galaxies, we broadened each spectrum to match the largest value of
$\sigma_\star{^2} + \sigma_{inst}{^2}$ of our sample, corresponding to
NGC 3115 ($\sigma_\star = 275$ and $\sigma_{inst} = 56$ km/s). This
correction changes the value of $W_{\rm CaT}$ by typically $10 \%$
with respect to the uncorrected values given in Paper I. $W_{\rm CaK}$
measurements from Cid Fernandes et al (2004) are also used in our
analysis.

\subsection{Seyfert 1s x Seyfert 2s}

\label{subsec:CaT in Sy}

It is well known that a non stellar light is present in Seyfert
1s. The contribution of such a component in Seyfert 2s has been a
topic of much controversy in the past. One way to address this issue
is to compare Seyfert 1s with Seyfert 2s through CaT.

\begin{figure}
\includegraphics[width=9cm]{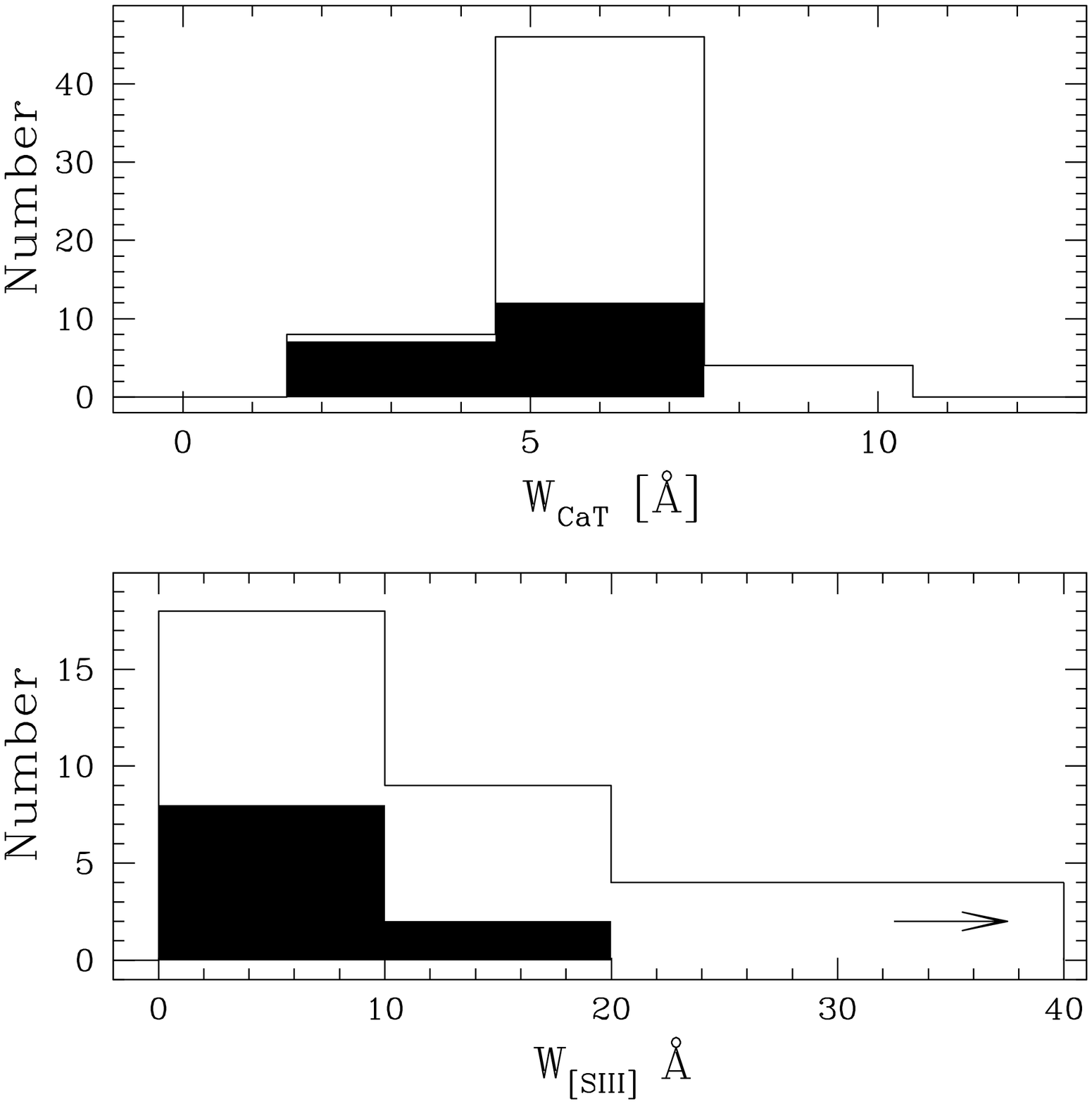}
\caption{Distributions of the equivalent width of the CaT absorption
lines (top) and of the [SIII]$\lambda9069$ emission line (bottom) for
Seyfert 1 (filled areas) and Seyfert 2 (empty areas).}
\label{Fig_hist_EWs}
\end{figure}

Fig. \ref{Fig_hist_EWs} (top panel) shows the distributions of $W_{\rm
  CaT}$ for Seyfert galaxies in our sample, with Seyfert 1s (including
subtypes 1--1.9) marked by the filled areas. Almost all Seyfert 2 lie
between 4 and 8 \AA, with a median of 6.5 \AA. $W_{\rm CaT}$ tends to
be smaller in type 1 nuclei, with a median of 4.8 \AA.  Emission lines
from the NLR follow the same pattern, with smaller equivalent widths
in Seyfert 1s than in Seyfert 2s, as shown for [SIII]$\lambda9069$ in
the bottom panel of Fig. \ref{Fig_hist_EWs}.

\begin{figure}
\includegraphics[width=9cm]{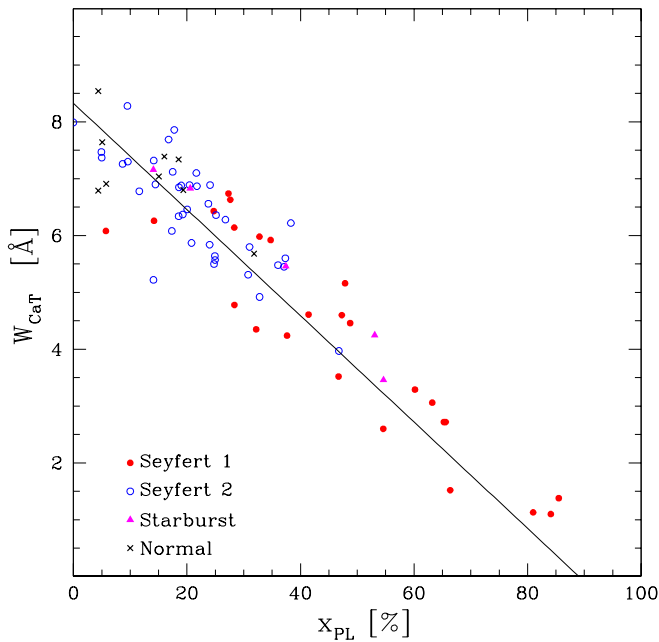}
\caption{$W_{\rm CaT}$ versus percentage of power-law contribution to
the synthesis. Hereafter, symbols are like in this figure.}
\label{Fig_CaT_x_PL}
\end{figure}

Another way of illustrating this result is shown in
Fig. \ref{Fig_CaT_x_PL}, where we plot the CaT strength as a function
the fractional contribution of power-laws to the continuum in the CaT
region. This fraction, which we denote $x_{PL}$, is deduced from the
spectral fits presented in Paper I, where, besides observed stars,
power-laws of several slopes were included to emulate the effects of a
template mismatch in the $\sigma_\star$ measurements. Table
\ref{NewCaT_and_PL} list $x_{PL}$, along with the new (corrected)
$W_{\rm CaT}$ values mentioned at the beginning of this section and
their errors.

\begin{table*}
\begin{center}
\begin{tabular}{lllrlllr}
\multicolumn{8}{c}{}\\
\hline
Name & Type & $W_{\rm CaT}$ (\AA) & $x_{PL}$ ($\%$) & Name & Type & $W_{\rm CaT}$ (\AA) & $x_{PL}$ ($\%$) \\ \hline
  NGC 205   &  Normal   &    5.68$\pm$0.46   & 32  &  NGC 7410        & Sy   2   &  7.10$\pm$0.46   &  22 \\
  NGC 526A  &  Sy 1.9   &    4.46$\pm$0.45   & 49  &  NGC 7469        & Sy 1.5   &  2.72$\pm$0.45   &  65 \\
  NGC 526B  &  Normal   &    6.80$\pm$0.47   & 19  &  NGC 7496        & Sy   2   &  5.45$\pm$0.47   &  37 \\
  NGC 1019  &  Sy 1.5   &    5.92$\pm$0.44   & 35  &  NGC 7582        & Sy   2   &  5.22$\pm$0.47   &  14 \\
  NGC 1068  &  Sy 2     &    5.84$\pm$0.48   & 24  &  NGC 7590        & Sy   2   &  7.12$\pm$0.43   &  17 \\
  NGC 1125  &  Sy 2     &    6.89$\pm$0.46   & 21  &  NGC 7714        & Sy   3   &  4.25$\pm$0.46   &  53 \\
  NGC 1140  &  Starburst&    5.46$\pm$0.44   & 37  &  MARK 0001       & Sy   2   &  5.64$\pm$0.46   &  25 \\
  NGC 1142  &  Sy 2     &    7.99$\pm$0.46   &  0  &  MARK 0003       & Sy   2   &  3.97$\pm$0.48   &  47 \\
  NGC 1241  &  Sy 2     &    7.86$\pm$0.43   & 18  &  MARK 0040       & Sy   1   &  4.24$\pm$0.46   &  38 \\
  NGC 1365  &  Sy 1.8   &    0.93$\pm$0.46   & 85  &  MARK 0078       & Sy   2   &  6.90$\pm$0.45   &  15 \\
  NGC 1380  &  Normal   &    7.64$\pm$0.46   &  5  &  MARK 0079       & Sy 1.2   &  2.64$\pm$0.46   &  55 \\
  NGC 1386  &  Sy 2     &    7.47$\pm$0.46   &  5  &  MARK 0273       & Sy   2   &  7.30$\pm$0.44   &  10 \\
  NGC 1433  &  Normal   &    6.79$\pm$0.44   &  4  &  MARK 0348       & Sy   2   &  5.87$\pm$0.44   &  21 \\
  NGC 1672  &  Starburst&    7.16$\pm$0.44   & 14  &  MARK 0372       & Sy 1.5   &  6.08$\pm$0.46   &   6 \\
  NGC 1808  &  Starburst&    6.83$\pm$0.46   & 21  &  MARK 0461       & Sy   2   &  5.57$\pm$0.46   &  25 \\
  NGC 2110  &  Sy 2     &    6.34$\pm$0.47   & 19  &  MARK 0516       & Sy 1.8   &  6.63$\pm$0.46   &  28 \\
  NGC 2639  &  Sy 1.9   &    6.74$\pm$0.44   & 27  &  MARK 0573       & Sy   2   &  7.26$\pm$0.44   &   9 \\
  NGC 2997  &  Normal   &    7.39$\pm$0.45   & 16  &  MARK 0705       & Sy 1.2   &  4.61$\pm$0.46   &  41 \\
  NGC 3081  &  Sy 2     &    6.89$\pm$0.46   & 24  &  MARK 0915       & Sy 1.8   &  4.78$\pm$0.47   &  28 \\
  NGC 3115  &  Normal   &    7.04$\pm$0.45   & 15  &  MARK 1066       & Sy   2   &  5.31$\pm$0.45   &  31 \\
  NGC 3256  &  Starburst&    3.46$\pm$0.46   & 55  &  MARK 1073       & Sy   2   &  5.50$\pm$0.46   &  25 \\
  NGC 3281  &  Sy 2     &    6.85$\pm$0.45   & 19  &  MARK 1210       & Sy   2   &  6.22$\pm$0.48   &  38 \\
  NGC 3783  &  Sy 1.5   &    2.72$\pm$0.45   & 66  &  MARK 1239       & Sy 1.5   &  0.92$\pm$0.47   &  81 \\
  NGC 4339  &  Normal   &    6.91$\pm$0.46   &  6  &  ESO 362-G08     & Sy   2   &  6.87$\pm$0.45   &  22 \\
  NGC 4507  &  Sy 2     &    6.56$\pm$0.45   & 24  &  ESO 362-G18     & Sy 1.5   &  6.14$\pm$0.45   &  28 \\
  NGC 4593  &  Sy 1     &    3.29$\pm$0.45   & 60  &  IC 2560         & Sy   2   &  7.32$\pm$0.45   &  14 \\
  NGC 4748  &  Sy 1     &    3.06$\pm$0.44   & 63  &  IC 3639         & Sy   2   &  5.80$\pm$0.46   &  31 \\
  NGC 4968  &  Sy 2     &    6.46$\pm$0.45   & 20  &  IC 5169         & Sy   2   &  6.88$\pm$0.45   &  19 \\
  NGC 5135  &  Sy 2     &    5.60$\pm$0.46   & 37  &  IRAS 01475-0740 & Sy   2   &  5.48$\pm$0.47   &  36 \\
  NGC 5929  &  Sy 2     &    6.08$\pm$0.46   & 17  &  IRAS 04502-0317 & Sy   2   &  6.37$\pm$0.47   &  19 \\
  NGC 6300  &  Sy 2     &    7.69$\pm$0.44   & 17  &  MCG -01-24-012  & Sy   2   &  6.28$\pm$0.47   &  27 \\
  NGC 6814  &  Sy 1.5   &    3.52$\pm$0.47   & 47  &  MCG -02-08-039  & Sy   2   &  7.37$\pm$0.46   &   5 \\
  NGC 6860  &  Sy 1.5   &    5.16$\pm$0.43   & 48  &  MCG -06-30-015  & Sy 1.5   &  4.60$\pm$0.45   &  47 \\
  NGC 6907  &  Normal   &    8.54$\pm$0.44   &  4  &  MCG +8-11-11    & Sy 1.5   &  0.92$\pm$0.44   &  84 \\
  NGC 6951  &  Sy 2     &    8.28$\pm$0.45   & 10  &  UGC 1395        & Sy 1.9   &  5.98$\pm$0.45   &  33 \\
  NGC 7130  &  Sy 1.9   &    6.43$\pm$0.45   & 25  &  UGC 12138       & Sy 1.8   &  6.26$\pm$0.44   &  14 \\
  NGC 7172  &  Sy 2     &    6.36$\pm$0.44   & 25  &  UGC 12348       & Sy   2   &  6.78$\pm$0.45   &  12 \\
  NGC 7184  &  Normal   &    7.34$\pm$0.44   & 19  &  UGC 3478        & Sy 1.2   &  1.28$\pm$0.44   &  66 \\
  NGC 7212  &  Sy 2     &    4.92$\pm$0.46   & 33  &  AKN 564         & Sy 1.8   &  3.05$\pm$0.49   &  32 \\
\end{tabular}
\end{center}
\caption{Results of $W_{\rm CaT}$ and $x_{PL}$. Columns lists galaxy
  name, type of activity, $W_{\rm CaT}$ and $x_{PL}$ contribution. All
  $W_{\rm CaT}$ data are measured in the same system, as explained in
  the text.}
\label{NewCaT_and_PL}
\end{table*}

The plot shows that $x_{PL}$ is smaller for Seyfert 2s than for
Seyfert 1s. Furthermore, the values of both $x_{PL}$ and $W_{\rm CaT}$
obtained for Seyfert 2s span approximately the same range as those
found for the non-active galaxies in the sample.

The fact that $W_{\rm CaT}({\rm Sey 1}) < W_{\rm CaT}({\rm Sey 2})$
and similarly $W_{[SIII]}({\rm Sey 1}) < W_{[SIII]}({\rm Sey 2})$ is
consistent with the unified model, which predicts a dilution of the
equivalent widths of the absorption and emission lines in sources
where the nuclear non-stellar continuum is seen directly. The fact
that the range of CaT strengths spanned by Seyfert 2s is similar to
that spanned by non-active galaxies indicates that their scattered FC,
if present, is too weak to produce a noticeable effect on the near-IR
range.

\subsection{Dilution by an underlying FC: The CaT versus CaK diagram}

\label{subsec:Dilution}

While the results reported above suggest one does not need to worry
about non-stellar light in the CaT range in Seyfert 2s, it is
desirable to test this idea more conclusively before proceeding to an
analysis of $W_{\rm CaT}$ in terms of stellar populations alone. This
can be done combining our CaT data with information from a different
wavelength range. The equivalent width of the CaII K absorption at
3933 \AA\ ($W_{\rm CaK}$) provides the necessary information.

The CaK line, which originates in old, late-type stars from the host's
bulge, is very sensitive to dilution by an underlying blue
continuum. As usual, this continuum can be either due to an AGN or to
a young stellar population, and the value of $W_{\rm CaK}$ by itself
cannot distinguish between these two possibilities (e.g.,
Storchi-Bergmann \etal 2000). The combination of $W_{\rm CaK}$ and
$W_{\rm CaT}$ data helps lifting this ambiguity.

The idea, first developed by TDT, is simple: if the CaK is diluted by
a truly featureless continuum, and if this FC extends from the optical
to the near IR, then the CaT would be diluted too, and by a
predictable amount. From observations of type 1 Seyferts and quasars,
we know that the AGN optical continuum is well described by a
power-law $F_\nu^{FC} \propto \nu^{-\alpha}$, with $\alpha$ in the
1--2 range or, equivalently, $F_\lambda^{FC} \propto
\lambda^{-\beta}$, with $\beta = 2 - \alpha$. This FC is superposed to
the bulge stellar light.

Suppose that the bulge has intrinsic CaK and CaT equivalent widths
$W_{\rm CaK}^\star$ and $W_{\rm CaT}^\star$, and that the ratio of the
stellar continuum fluxes over these two lines is $c^\star \equiv
C_{\lambda_{\rm CaK}}^\star / C_{\lambda_{\rm CaT}}^\star$.  Trivial
algebra then leads to the following relation between the observed CaK
and CaT equivalent widths:

\begin{equation}
W_{CaK} = W_{CaK}^* (1-x_{CaK}^{FC}).
\end{equation}

\begin{equation} \label{eq:dilution} W_{CaT} = W_{CaT}^*
\frac{1}{1+{c^\star \times
(\frac{\lambda_{CaT}}{\lambda_{CaK}})^{-\beta} \times
\frac{x_{CaK}^{FC}}{1-x_{CaK}^{FC}}}} \end{equation}

\noindent where $x^{\rm FC}_{CaK}$ is the fraction of the total
continuum at 3933 \AA\ which comes from the FC.

We modeled a mixing of a FC and bulge continuum in a form of ``FC +
SSP'' mixing lines, where SSP refers to a Simple Stellar Population
representative of 10 Gyr old, typical of elliptical galaxies and
bulges, as obtained from Bruzual \& Charlot (2003, BC03) models for
solar metallicity. Practical use of equation \ref{eq:dilution}
requires stipulating fiducial values for $W_{\rm CaK}^\star$, $W_{\rm
  CaT}^\star$ and $c^\star$. We adopt $(W_{\rm CaK}^\star,W_{\rm
  CaT}^\star) = (21.1,7.9)$ \AA\ and $c^\star = 0.38$.

\begin{figure}
\includegraphics[width=9cm]{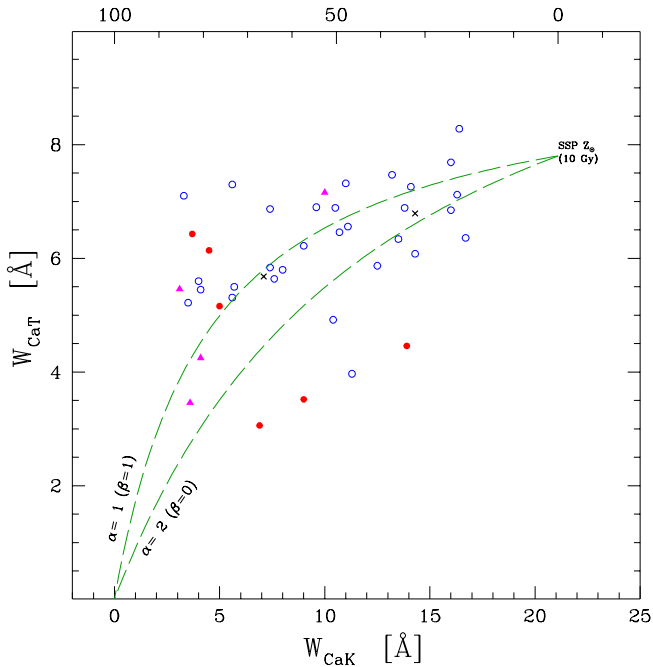}
\caption{Equivalent width of the CaT (from Paper I) plotted against
  the equivalent width of the CaII K line (from the literature). The
  lines show the expected dilution curves (equation \ref{eq:dilution})
  for bulge plus FC power-laws, with slopes $\alpha$ = 1 and 2
  (typical range of type 1 AGN), as labeled. Upper scale shows the
  percentage of $x_{CaK}^{FC}$ contribution. Symbols as in
  Fig. \ref{Fig_CaT_x_PL}.}
\label{Fig_CaT_x_CaK_a}
\end{figure}

Equation \ref{eq:dilution} is over plotted to the data points in the
$W_{\rm CaT}$ versus $W_{\rm CaK}$ diagram
(Fig. \ref{Fig_CaT_x_CaK_a}). The lines indicate the bulge + FC mixing
lines for different values of the slope $\alpha$ and for
$x_{CaK}^{FC}$ decreasing from 100\% at $(W_{\rm CaT},W_{\rm CaK}) =
(0,0)$ to 0\% at $(W_{\rm CaT},W_{\rm CaK}) = (W_{\rm
  CaT}^\star,W_{\rm CaK}^\star)$. The $W_{\rm CaK}$ measurements come
from the optical studies of Cid Fernandes \etal (1998, 2001, 2004). Of
the 78 galaxies in our sample, 42 overlap with these studies, most of
which (31) correspond to Seyfert 2s.

We see that some Seyfert nuclei line up along the mixing
lines. However, in order to explain all of the objects, it is
necessary a contribution of $x_{CaK}^{FC}$ of about 50\% or
greater. There are some Seyfert with high $W_{\rm CaT}$ and low
$W_{\rm CaK}$ that are not explained in terms of these mixing lines
for $\alpha$ between 1 ($\beta$=1) and 2 ($\beta$=0), which are
representative of type 1 AGN. For these objects, only if $\alpha$ were
as low as $\sim 0.5$ or less the dilution lines would go approximately
through the observed values, but not even quasars have such an
extremely blue continuum. Such a blue FC could in principle be
produced by scattering of the AGN light by dust
particles. Nevertheless, the required values of the $x_{CaK}^{FC}$
fraction (of the order of $\sim 50 \%$) would imply that scattered
broad lines in the optical should become visible, in which case the
galaxy would not be classified as a Seyfert 2 in the first place (an
argument first put forward by Cid Fernandes \& Terlevich
1995). Besides that, about half of the Seyfert 2s with $x_{CaK}^{FC}$
greater than 50\% show High Order Balmer Lines (HOBL) in absorption in
their spectra (Gonz\'alez Delgado \etal 2001). So, not diluted HOBL
argue against a dilution by a power-law in these objects.

We have to point out that ``FC + SSP'' in Fig. \ref{Fig_CaT_x_CaK_a}
are computed with a solar SSP values. For another metallicities,
fidutial numbers are a litlle different (mainly lower $W_{\rm CaT}$),
and the lines would go down for about 1 \AA. In that way, mixing lines
in Fig. \ref{Fig_CaT_x_CaK_a} should be considered as an upper limit
for ``FC + SSP'' models.

We also computed ``FC + E'' models (not shown in
Fig. \ref{Fig_CaT_x_CaK_a}), where E refers to NGC 2110, the earliest
(t=-3) Seyfert 2 galaxy in our sample for which we have determined
$W_{\rm CaK}$ and $W_{\rm CaT}$. For this model, $(W_{\rm
  CaK}^\star,W_{\rm CaT}^\star) = (13.5,6.34)$ \AA, and the dilution
lines would stay even more distant from the data for the same range of
$\alpha$ considered before. Not even for $\alpha$ as low as $\sim$ 0
(not realistic) the models would along up the data. So, our main
conclusions about FC + bulge mixing models for Seyfert 2 nuclei
remains valid.

TDT have performed the same analysis, but using the MgIb line at 5175
\AA\ instead of CaK (see also Jim\'enez-Benito \etal 2000). They
conclude that if the observed dilution of MgIb is due to a power-law
FC of reasonable slope, then the CaT would also be diluted, contrary
to their empirical result. Here, we considered ``FC + SSP'' mixing
models with some reasonable combination of $W_{\rm CaK}^\star$,
$W_{\rm CaT}^\star$ and $c^\star$ values representative of old stellar
populations. These models shows that dilution by a power-law FC seems
to be not completely adequate to account for the observed strengths of
optical and near-infrared ranges simultaneously. Hence, a simple
mixture of a bulge-like component plus an AGN-like FC, envisaged in
the early days of AGN research (e.g., Koski 1978), is just not a
viable description of the data. We will analyse another explanation
for $W_{\rm CaK}$ and $W_{\rm CaT}$ behavior in the next section.

\section{CaT strength as a Stellar Population diagnostic}
\label{Stellar Models}

Now that we considered dilution by non-stellar light in the CaT range
in type 2 Seyferts, we will proceed to a stellar population analysis.

The behavior of the CaT equivalent width as a function of stellar
parameters ($T_{\rm eff}$, $\log g$, [Fe/H]) has been addressed by
several studies (e.g., D\'iaz, Terlevich \& Terlevich 1989), and this
knowledge has been incorporated into population synthesis models which
map the CaT dependence on the age and metallicity of coeval stellar
populations (e.g., Idiart, Thevenin \& de Freitas Pacheco 1997). In a
recent and comprehensive study, Cenarro and coworkers revisited this
topic, analyzing the CaT in a large library of stars and producing
state-of-the-art predictions for the behavior of the CaT in integrated
stellar populations (Cenarro \etal 2001a,b,2002; Vazdekis \etal 2003).

Despite early hopes that the CaT could serve as a metallicity
indicator for galaxies (as it does for globular clusters, at least
below $\sim 0.3 Z_\odot$; Armandroff \& Zinn 1988; Idiart \etal 1997;
Vazdekis \etal 2003), or as tracers of recent star-formation through
its sensitivity to the presence of red super-giants (hereafter RSG;
see TDT; Garc\'{\i}a-Vargas, Moll\'a \& Bressan 1998; Mayya 1997),
observations have revealed that the CaT shows remarkably little
variation in strength for galaxies spanning wide ranges in
morphological and stellar population properties. This small
sensitivity is qualitatively consistent with the new generation of
synthesis models (Vazdekis \etal 2003), which show that, for
metallicities higher than $\sim -0.5$ dex, this feature varies little
as a function of age and metallicity.

Overall, these results raise serious doubts as to the usefulness of
the CaT as a stellar population diagnostic.  In this section we take
another look at this issue, tackling it from an empirical perspective.
As in the previous section, information on the CaK line will be
incorporated in the analysis.

\subsection{CaT versus CaK}
\label{subsec:CaT_x_CaK}

Let us assume that strongly diluted CaK lines in Seyfert 2s are the
result of the presence of a Starburst component, in line with previous
investigations. In particular, Cid Fernandes \etal (2001) propose that
Seyfert 2s with $W_{\rm CaK} \le 10$ \AA\ can be safely identified as
composite Starburst + AGN systems, where the Starburst is
unambiguously identified by independent features such as the WR bump
or PCygni line profiles in the UV. Investigating the behavior of
$W_{\rm CaT}$ as a function of $W_{\rm CaK}$ thus provides a fully
empirical test of whether the CaT is or is not sensitive to stellar
populations.

A CaT versus CaK diagram was already shown in
Fig.\ \ref{Fig_CaT_x_CaK_a}, where most points correspond to Seyfert
2s. The plot shows that $W_{\rm CaT}$ bears little, if any, relation
to $W_{\rm CaK}$. It is therefore hard to evade the conclusion that
for the mean ages and metallicities of metal rich stellar populations,
$W_{\rm CaT}$ is a poor diagnostic: objects as diverse as Mrk 1210,
which contains a powerful young Starburst, including Wolf-Rayet stars
(Storchi-Bergmann, Cid Fernandes \& Schmitt 1998), and NGC 7172, which
is dominated by $\sim 10$ Gyr stars (Cid Fernandes et al 2004) have
$W_{\rm CaT}$ values indistinguishable within the errors ($6.7 \pm
0.4$ and $6.9 \pm 1.1$ \AA, respectively).

\begin{figure}
\includegraphics[width=9cm]{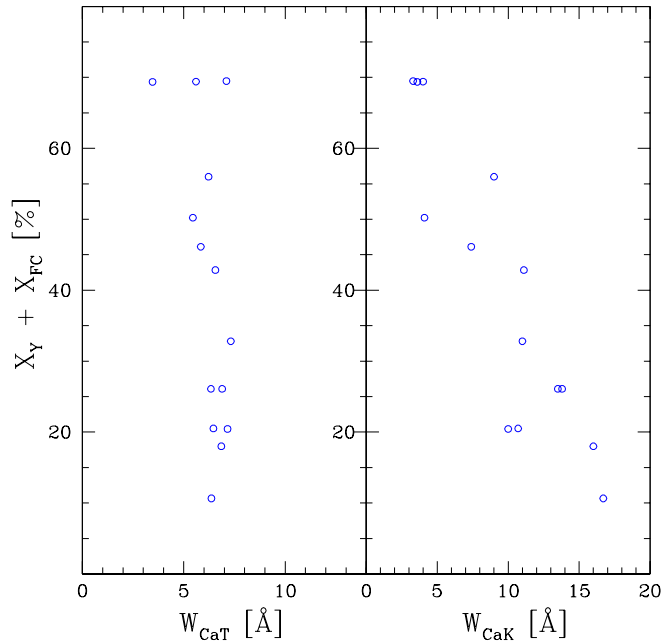}
\caption{Percentage of a Starburst component (see text) contribution
to the light in synthetic spectra versus $W_{\rm CaT}$ (left) and
$W_{\rm CaK}$ (right) in Seyfert 2s. Note the contrast of the
behavior of $W_{\rm CaK}$, which is very sensititive to young stars,
with the approximate constancy of $W_{\rm CaT}$.}
\label{Fig_Perc_x_EWs}
\end{figure}

In Fig.\ \ref{Fig_Perc_x_EWs} we combine our CaT data with the results
of the detailed spectral synthesis analysis of type 2 Seyferts carried
out by Cid Fernandes \etal (2004) for 14 sources in common with our
sample. We compare their young stars (age $< 25$ Myr) plus ``FC''
light fractions at 4020 \AA\ ($x_{Y/FC}$) with both $W_{\rm CaT}$
(left panel), and $W_{\rm CaK}$ (right). As discussed in that paper,
which dealt exclusively with spectra in the 3500--5200 \AA\ range,
what the synthesis returns as a FC is in most cases a reddened
Starburst component instead of a true non-stellar component, such that
$x_{Y/FC}$ is in practice a measure of the Starburst strength.

The near constancy of $W_{\rm CaT}$, already evident in
Fig.\ \ref{Fig_CaT_x_CaK_a}, is even more striking in
Fig. \ref{Fig_Perc_x_EWs}. The inexistent correlation between CaT
strength and $x_{Y/FC}$ contrasts with the strong decrease in $W_{\rm
  CaK}$ as the Starburst component increases in strength. Again, one
is lead to the conclusion that, by itself, $W_{\rm CaT}$ is not a
stellar population tracer for our galaxies.

\subsection{Evolutionary Synthesis Models in the CaT versus CaK plane}

\label{subsec:BC03}

The most natural interpretation of Figs. \ref{Fig_CaT_x_CaK_a} and
\ref{Fig_Perc_x_EWs} is that the dilution of optical lines in some
Seyfert 2 nuclei is due to a young stellar population which causes
little or no dilution of the CaT. In this section we use models to
check whether this is a viable scenario.

To investigate the behavior of the CaT and CaK lines as a function of
stellar population properties, we have computed the evolution of
$W_{\rm CaT}$ and $W_{\rm CaK}$ for SSP for 4 metallicities. BC03
evolutionary synthesis models were used for this purpose. SSP models
are computed with the STELIB library (Le Borgne \etal 2003), Chabrier
(2003) mass function and Padova 1994 tracks (see BC03 for
details). Because $W_{\rm CaT}$, $W_{\rm CaT*}$ and $W_{\rm CaK}$ are
not provided in the standard BC03 distribution, they were computed
directly from the theoretical spectral energy distributions, as
explained at the beginning of Section
\ref{sec:CaT_NonStellarLight}. For compatibility with our $W_{\rm
  CaT}$ measurements, we broadened all model spectra as explained at
the beginning of section \ref{sec:CaT_NonStellarLight}, adopting for
the instrumental resolution that of the STELIB library ($\sim 45$ km/s
at $\sim 8600$ \AA).

\begin{figure}
\includegraphics[width=9cm]{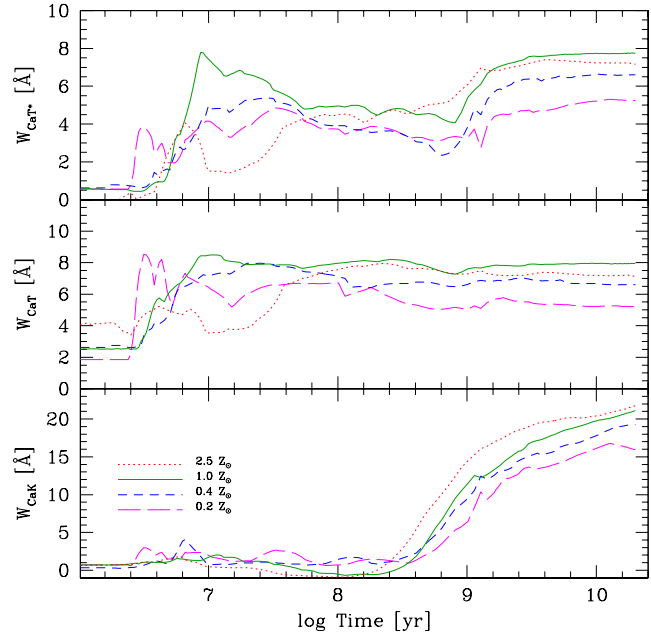}
\caption{Evolution of the $W_{\rm CaT*}$, $W_{\rm CaT}$ and $W_{\rm
    CaK}$ absorption line indices for SSP, from BC03 models.  Long
  dashed, dashed, solid and dotted lines correspond to metallicities
  of 0.2, 0.4, 1 and 2.5 $Z_\odot$, respectively.}
\label{Fig_BC03_x_t_a}
\end{figure}

Fig. \ref{Fig_BC03_x_t_a} shows the results for four different
metallicities: 0.2, 0.4, 1 and 2.5 $Z_\odot$. $W_{\rm CaK}$ is
practically null during the first $\sim 10^{8.5}$ yr. After these
initial phases, the CaK strength increases monotonically with time,
which makes it good to discriminate very young SSP (younger than $\sim
3 \times 10^8$ yr) from SSP older than $\sim 1$ Gyr. On the other
hand, $W_{\rm CaT}$ starts from small values up to a few Myr, when the
first RSG appear. The exact timing and duration of this phase depends
on the metallicity. From $\sim 10$ Myr to about 1 Gyr, $W_{\rm CaT}$
oscillates as RSG are replaced by red-giants. Throughout these
different phases, however, $W_{\rm CaT}$ spans a small range, from
$\sim 6$ to 8 \AA.  Taking into account the uncertainties in both
models and data, this is in practice a constant value. Only in the
first Myrs of evolution does $W_{\rm CaT}$ have a useful diagnostic
power, as it assumes much smaller values. Notice that in these early
phases $W_{\rm CaT}$ is contaminated by Paschen lines, which explains
why it does not start from 0 at $t = 0$.

In Fig. \ref{Fig_CaT_x_CaK_b} these models are overlayed to the data
in the $W_{\rm CaT}$ versus $W_{\rm CaK}$ diagram. Seyfert 1s are not
plotted because of their non-stellar continuum, which affects both CaK
and CaT. Numbers along the model tracks mark ages of $10^6$, $10^7$,
$10^8$, $10^9$ and $10^{10}$ yr. The first thing to notice is that,
unlike the bulge + FC mixing lines in Fig. \ref{Fig_CaT_x_CaK_a},
these models do span the region occupied by objects in our sample,
showing that stellar populations alone are capable of explaining the
behavior of Seyfert 2s in the CaK versus CaT diagram. Secondly,
Fig. \ref{Fig_CaT_x_CaK_b} seems to indicate that sub-solar abundances
may be required to match the data on Seyfert 2s, particularly those
with weak CaT. This is somewhat surprising, given that we are dealing
with the nuclei of early type spirals, where $Z$ is expected to be
relatively high. One must nevertheless recall that these are SSP
models, whereas actual galaxies contain a mixture of stellar
populations. This is particularly true for type 2 Seyferts, where
previous studies have revealed a wide variety of star formation
histories (Cid Fernandes \etal 2001; 2004; Wild \etal 2007). More
realistic star formation histories can be modeled as a mixture of SSP,
whose effect on this diagram can be conceived drawing imaginary mixing
lines connecting any series of points along the instantaneous burst
tracks.

\subsection{Mixed models: Young plus Old populations}
\label{Mixed Models}

To have a rough idea of the effects of such a mixture, in
Fig. \ref{Fig_CaT_x_CaK_c} we repeat the spectral mixing experiment of
Fig. \ref{Fig_CaT_x_CaK_a} but this time using a 1 Myr population (to
represent ongoing star-formation) instead of a non stellar power-law
continuum. Solar and 2.5 solar BC03 models are used. We further allow
for the possibility that the young population suffers more extinction
than the old one, so we consider three cases, $A_{V}$ = 0, 1.5 and 3
mag. Again, we do not plot Seyfert 1 nuclei, as they present dilution
due to the non-stellar continuum and thus are out of the scope of this
experiment. This exercise is essentially the same as that performed
with a power-law FC in Fig. \ref{Fig_CaT_x_CaK_a}, except that the
continuum shape is not the same, and, due to contamination by Paschen
lines, $W_{\rm CaT}$ is not $= 0$ for a pure Starburst (see
Fig. \ref{Fig_BC03_x_t_a}, middle panel).

As is evident in the Figure, these mixing lines span the observed
points for most of the Sy 2 sample. It is important to emphasize that
only for very young populations one expects a relatively low CaT. As
shown in the plot, galaxies with $W_{\rm CaT} < 6$ \AA\ and $W_{\rm
  CaK} < 10$ \AA\ require such a population to be adequately modeled.
Not even the $0.4$ $Z_\odot$ models are capable of modeling these
objects without a very young population.

It is also important to note that although our 1 Myr + 10 Gyr mixing
models span most of the observed points in the CaK versus CaT plane,
galaxies with $W_{\rm CaT}$ above $\sim 6$ \AA\ can be equally well
modeled with an older young population.  For metallicites $\ge 0.4$
$Z_\odot$, $W_{\rm CaT}$ rises above 6 \AA\ as early on as $\sim 5$
Myr, with the appearence of the first RSGs. This also happens for
continuous star formation models, as shown by the solid line in the
CaT-CaK diagram.

\begin{figure}
\includegraphics[width=9cm]{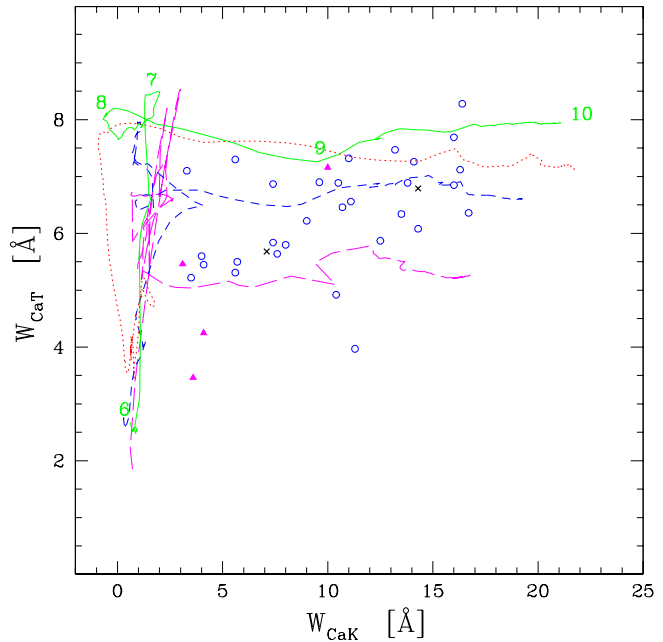}
\caption{Same as Fig. \ref{Fig_CaT_x_CaK_a} but without Seyfert 1
  nuclei and over plotting simple stellar populations, as computed
  from the BC03 models. Lines as in Fig. \ref{Fig_BC03_x_t_a}. Numbers
  in stellar lines denotes log(Age[yr]). Symbols as in
  Fig. \ref{Fig_CaT_x_PL}.}
\label{Fig_CaT_x_CaK_b}
\end{figure}

\begin{figure}
\includegraphics[width=9cm]{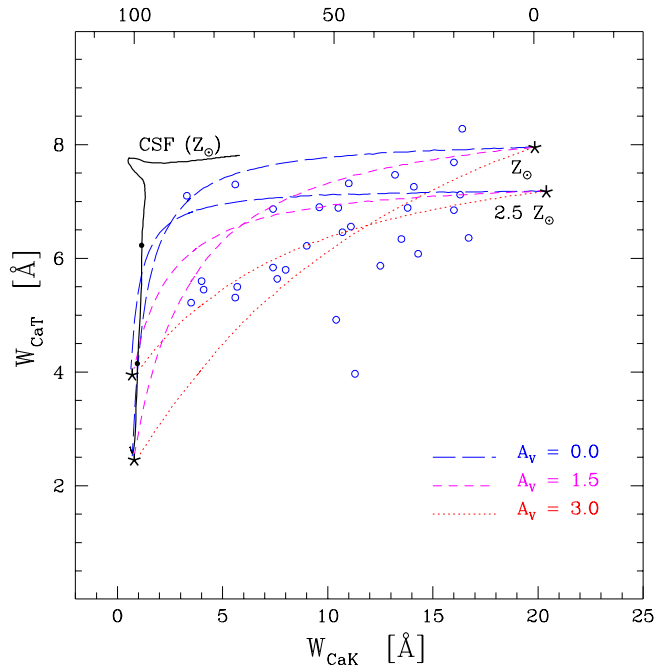}
\caption{Same as Fig. \ref{Fig_CaT_x_CaK_b} now overploting old plus
  redenned young stellar population models for solar and 2.5 solar
  metallicity. Upper scale shows the percentage of the Starburst
  contribution in the CaK region, $x_{CaK}^{Stb}$. Mixture lines
  corresponds to different extinction magnitudes and the asterisks
  denote null ($0 \%$, upper right) and full ($100 \%$, lower left)
  young contribution. The CSF line refers to continuous star formation
  from ages between 1 Myr and 10 Gyr. Dots on this line indicates
  bursts of 5 and 10 Myr old.}
\label{Fig_CaT_x_CaK_c}
\end{figure}

Mixing models for $\ge 5$ Myr + 10 Gyr populations are not shown for
clarity, but, as can be deduced from Fig. \ref{Fig_CaT_x_CaK_b}, they
would span most of the upper part of CaK versus CaT plane, where many
Sy 2s are located. We note in passing that TDT originally suggested
that the CaT in Seyfert 2s signal the presence of RSG in a
Starburst. While this may well be true in some cases, our modeling
shows that there is no way of telling it just from the strength of the
CaT, since we are able to fit the $W_{\rm CaT}$ and $W_{\rm CaK}$ data
both with and without RSGs.

Given the simplicity of these models, the overlap between data and
models in the CaT-CaK diagram is highly encouraging. Nevertheless, at
the middle of the diagram there are two Sy 2 that are far away from
our models. The two ``outliers'' are Mrk 3 and NGC 7212. It may be
that a bona-fide AGN continuum is present in these galaxies, which
would help explaining their low CaK and specially CaT lines.  The
spectral synthesis analysis of NGC 7212 by Cid Fernandes \etal (2004)
indicates that $\sim 35\%$ of the light at 4020 \AA\ comes from an FC
component.  They also identify a weak broad H$\beta$, strengthening
the case for the presence of a non-stellar continuum. Furthermore, for
both NGC 7212 and Mrk 3, imaging (Pogge \& De Robertis 1993;
Kotilainen \& Ward 1997), spectropolarimetry (Tran 1995), and spectral
modelling (Gonz\'alez Delgado \etal 2001) all point to the existence
of an FC component (see Cid Fernandes \etal 2001 for a detalied
discussion).  None of these studies find compelling evidence for
significant on-going star-formation in these galaxies. It is therefore
likely that their low $W_{\rm CaT}$ values (the smallest amongst
Seyfert 2s) are indeed due to dilution by non-stellar light,
explaining why they deviate from the region spanned by our purely
stellar models in Fig. \ref{Fig_CaT_x_CaK_c}.

A consistent scenario thus emerges, where the blue continuum which
dilutes the optical spectrum of some Seyfert 2s is due to young
stellar populations, while their undiluted CaT strengths can be
explained as due to RSGs or older populations. As explained above, in
the few cases where the CaT is low, a combination of an old plus a
very young (pre-RSG) population is needed. In these cases, one may
expect to see Paschen lines in absorption, which is not observed in
our galaxies, as reported in Paper I.

To investigate this potential problem we have examined the strength of
Paschen lines in the theoretical spectra of the mixing models shown in
Fig. \ref{Fig_CaT_x_CaK_c}. We pay particular attention to the Pa14
line at $8598$ \AA, which is strategically placed between Ca2 and
Ca3. We find that, for $A_V = 0$, Pa14 only becomes visible for
$x^{Stb}_{\rm CaK} > 95 \%$ or $85 \%$ for Z = 1 and 2.5 $Z_{\odot}$,
respectively. These fractions falls to about 75 $\%$ is we consider 3
magnitudes of differential extinction. These very large fractions
result from the fact that the young population is much bluer than the
old one, so any mixture where young stars contributes significantly to
the CaT range, automaticaly implies that this same component
completely dominates in the optical. Spectral synthesis studies in the
optical range never find such huge starburst contribution in Seyfert
2s. Thus, the non detection of Paschen lines may be simply due to the
fact that old stars dominate the flux in the near IR, even when
younger populations dominate the optical light.

A further factor which may be related to the non detection of Paschen
absorption is filling by an emission component, powered either by the
AGN or a circumnuclear starburst. A few of our galaxies do show a hint
of Pa14 emission, like the Starburst galaxy NGC 7714, where Pa14 has
about 0.5 \AA\ emission equivalent width (see also Gonz\'alez Delgado
\etal 1995). In other cases, emission and absorption may approximately
compensate each other. This seems to be the case in the super star
clusters of NGC 1569, where WR features, high-order Balmer absorption
lines and the CaT are all clearly detected (Gonz\'alez Delgado \etal
1997), but Paschen lines are not seen either in absorption or
emission.  Hence, even though further studies concerning Paschen lines
are desirable, their no detection do not invalidate our conclusions.

\section{CaT strength as a function of distance to the nucleus}

\label{subsec:Spatial CaT Profiles}

All CaT data reported so far pertains to nuclear regions. For 34 of
the 78 galaxies studied in Paper I our long-slit data are of
sufficient quality to warrant the extraction of spatially resolved
spectra. Detailed results for all such sources are presented in Asari
(2006). In this section we summarize the main results concerning the
spatial behavior of the CaT strength.

\begin{figure}
\includegraphics[width=0.5\textwidth, bb=50 180 420
575]{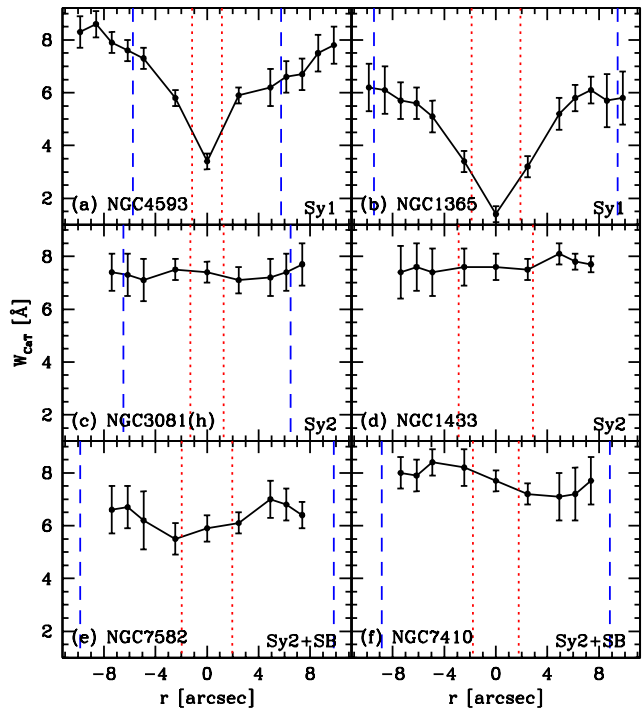}
\caption{Examples of $W_{\rm CaT}$ spatial profiles. Dotted lines mark
  $-200$ to +200 pc region; dashed lines, the $-1$ to +1 kpc
  region. Panels (a) and (b) show the dilution of $W_{\rm CaT}$ in
  Seyfert 1; panels (c)--(f) show how flat $W_{\rm CaT}$ is for
  Seyfert 2 and Seyfert 2 + Starburst.}
\label{fig:Spatial_WCaT}
\end{figure}

Fig. \ref{fig:Spatial_WCaT} shows $W_{\rm CaT}$ spatial profiles for 6
galaxies, chosen to represent Seyfert 1s (panels a and b), Seyfert 2s
(c and d) and composite Starburst + Seyfert 2 (e and f). There is a
clear drop in $W_{\rm CaT}$ as one approaches the nucleus in Seyfert
1s, whereas Seyfert 2s present remarkably flat $W_{\rm CaT}(r)$
profiles. At a few hundred pc from the nucleus, however, the CaT
strength becomes similar for both Seyfert types. This illustrates once
again how sensitive $W_{\rm CaT}$ is to an underlying FC, in agreement
with the results reported in Section \ref{sec:CaT_NonStellarLight}.

\begin{figure}
\includegraphics[width=9cm]{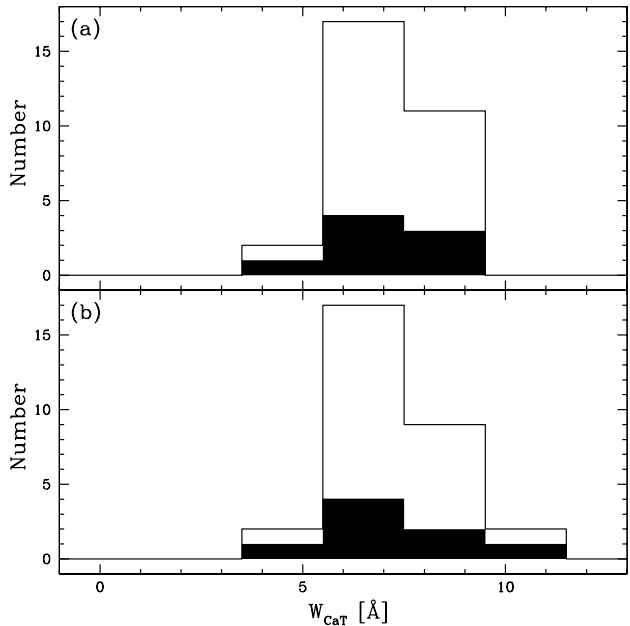}
\caption{Similar to Fig. \ref{Fig_hist_EWs} but for off-nuclear
  measurements: (a) 1 kpc data, (b) furthest point for which we have
  signal (see text for details). Scale and areas as in Fig
  \ref{Fig_hist_EWs}.}
\label{Fig_hist_EWs_Spatial}
\end{figure}

In a way, these results are arguably more convincing proof of the FC
effects.  Since the analysis in Section \ref{sec:CaT_NonStellarLight}
was based on a comparison of CaT strengths for Seyfert 1s and 2s, or
with fiducial values, one could worry that for some reason the stellar
populations of the two kinds of AGN are intrinsically different,
invalidating a comparative analysis. Here, on the other hand, we show
that at least immediately outside their nuclear regions, Seyfert 1s
and 2s are very similar, and that dilution is really a nuclear effect
present only in type 1s. To further illustrate this,
Fig. \ref{Fig_hist_EWs_Spatial} repeats the comparison performed in
Fig. \ref{Fig_hist_EWs}, but this time using off-nuclear $W_{\rm CaT}$
measurements. Depending of the galaxy redshift, the furthest
off-nuclear measurements correspond to less than 1 kpc to $\sim$ 3
kpc. In Fig. \ref{Fig_hist_EWs_Spatial}a we proceed as follows: in the
first cases (nearest galaxies) we take the furthest points from the
nucleus (for instance, NGC 1433 in Fig. \ref{fig:Spatial_WCaT}), and
in the second cases (furthest galaxies), we firstly obtained the
interpolated CaT values at exactly 1 kpc taking into account the data
at each side of the nucleus, and then we calculate the average of the
interpolated values (for instance, NGC 3081 in Fig
\ref{fig:Spatial_WCaT}). We see that the differences between Seyfert
1s and 2s disappear when considering these off-nuclear values, which
we call ``{\it $\le$1kpc}'' data. Fig. \ref{Fig_hist_EWs_Spatial}b
does the same thing, but using the values corresponding to the
furthest points from the nucleus for which we have signal, and we call
them ``{\it furthest}'' data. Numerical results of $W_{\rm CaT}$
spatial measurements are summarized in Table \ref{TableSpatialCat}. A
KS test applied to the distributions yields probablitities of $p=0.46$
(``{\it $\le$1kpc}'') and $p=0.84$ (``{\it furthest}''). We see that
even for short distances ($\le$ 1 kpc) there is no significant
dilution of CaT in Seyfert 1s, as we expected from the regions outside
the nuclei.

\begin{table}
\begin{center}
\begin{tabular}{lccc}
\\ \hline
                 & Nuclear [\AA] & $\le$1 kpc [\AA] & Furthest [\AA] \\ \hline
Seyfert 1 -- min &     2.7 &       4.3 &       5.0 \\
Seyfert 1 -- med &     4.8 &       6.5 &       6.6 \\
Seyfert 1 -- max &     6.7 &       8.3 &       9.0 \\
\hline
Seyfert 2 -- min &     4.0 &       3.9 &       3.7 \\
Seyfert 2 -- med &     6.5 &       6.4 &       6.4 \\
Seyfert 2 -- max &     8.3 &       8.3 &       9.1 \\
\hline
\end{tabular}
\end{center}
\caption{Statistics for Spatial $W_{\rm CaT}$ measurements: minimum,
  median and maximum values are shown for Seyfert 1 and Seyfert 2 and
  for nuclear and off-nuclear data. For off-nuclear data, we
  considered {\it $\le$1 kpc} data, which refers to the values at 1
  kpc or less from the nucleus, and {\it furthest} data, which are the
  corresponding values of furthest points from the nucleus of each
  galaxy for which we have signal.}
\label{TableSpatialCat}
\end{table}

It is also interesting to point out that these results are essentially
identical to those obtained from the spatial analysis of the CaK
strength by Cid Fernandes \etal (1998) and Storchi-Bergmann \etal
(1998, their Fig. 1). Hence, the FC which dilutes the optical spectrum
in Seyfert 1s is the same which decreases the CaT strength in the near
IR. This component is not present in type 2s.

The behavior of $W_{\rm CaT}(r)$ for Starburst $+$ Seyfert 2
composites (Figs.\ \ref{fig:Spatial_WCaT}e and f), on the other hand,
contrasts with that derived from $W_{\rm CaK}$ radial
profiles. Contrary to the strong spatial dilution by young stars
detected in the CaK and other optical lines, the CaT profiles show
very little, if any, variation as a function of distance to the
nucleus in composite systems. In NGC 7582 there is a marginal hint of
dilution. From optical work we know that this galaxy hosts a very
dusty central Starburst, so it could represent a case where the
Starburst light does contribute to the flux in the CaT range with a
non negligible fraction. However, even in this most favorable case the
nuclear CaT is diluted by less than 15\% with respect to its value at
$r \sim \pm 500$ pc.  In contrast, the optical absorption lines are
much more diluted: At the wavelengths of the CaK, G-band and MgIb
lines, the dilution factors are 70, 60 and 57\%, respectively.  To
better appreciate this strong difference, we invite the reader to
compare Fig.\ \ref{fig:Spatial_WCaT}e with fig. 29 of Cid Fernandes
\etal (1998), where the radial profiles of several optical absorption
lines in NGC 7582 are shown.

The difference in $W_{\rm CaT}$ and $W_{\rm CaK}$ profiles is
striking, but not surprising. In fact, our finding that the CaT does
not mimic the CaK spatial behavior is fully consistent with the
conclusion laid out in the section above that only under extreme
conditions (very young age and large extinction) would a Starburst
strongly dilute both CaK and CaT lines.

\section{Kinematics}

\label{sec:Kinematics}

Our spectroscopic data (Paper I) provides estimates of two
characteristic velocities: the typical velocity of stars through the
host's bulge, $\sigma_\star$, and the typical NLR cloud velocity
dispersions, inferred from the width of the [SIII]$\lambda$9069 line:
We fitted the profile of [SIII]$\lambda$9069 with one or two
Gaussians, one representing the core of the emission line and the
other, if necessary, representing the wings. The resulting FWHM of the
core component divided by $(8 \ln2)^{1/2}$ is what we call
$\sigma_{\rm [SIII]}$. We have determined $\sigma_{\rm [SIII]}$ by
this method for 31 Seyfert and 4 Starburst nuclei.

\begin{figure}
\includegraphics[width=9cm]{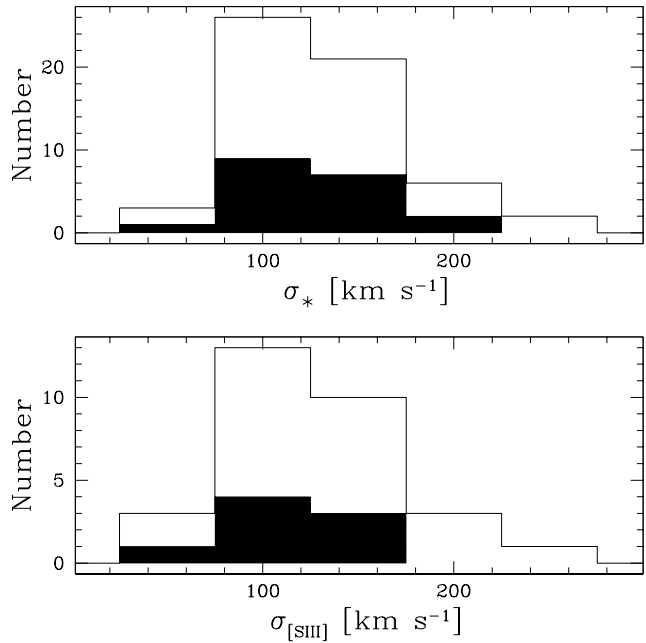}
\caption{Distributions of stellar velocity dispersions
  ($\sigma_{\star}$, top) and [SIII]-based gas velocity dispersions
  ($\sigma_{\rm [SIII]}$, bottom). Areas as explained in
  Fig. \ref{Fig_hist_EWs}.}
\label{Fig_histograms_VD}
\end{figure}

The distributions obtained for $\sigma_\star$ and $\sigma_{\rm
  [SIII]}$ are shown in Fig. \ref{Fig_histograms_VD}. They show that
both types of nuclei have essentially the same (statistical) values:
for Seyfert 1s $\sigma_\star$ = 128 $\pm$ 37 km/s and $\sigma_{\rm
  [SIII]}$ = 111 $\pm$ 29 km/s, while for Seyfert 2s $\sigma_\star$ =
134 $\pm$ 45 km/s and $\sigma_{\rm [SIII]}$ = 137 $\pm$ 76
km/s. Firstly, we will discuss the stellar velocity dispersions
($\sigma_\star$) and then we will analyze the NLR kinematics through
$\sigma_{\rm [SIII]}$ and its possible link to $\sigma_\star$.

\subsection{$\sigma_\star$ versus $W_{\rm CaT}$}

\label{subsec:Stellar velocity dispersions}

As reviewed in Section \ref{sec:Introduction}, recent work on normal
galaxies has dedicated much attention to the fact that $W_{\rm CaT}$
is slightly anti-correlated with velocity dispersion all the way from
dwarf to giant elliptical galaxies and bulges (Saglia \etal 2002;
Cenarro \etal 2003; Falc\'on-Barroso \etal 2003; Michielsen \etal
2003). Even though our CaT data are not of the same quality as that
employed in these studies, it is interesting to check whether
Starburst and active galaxies follow the same $W_{\rm
  CaT}$-$\sigma_\star$ as normal ones.

\begin{figure}
\includegraphics[width=9cm]{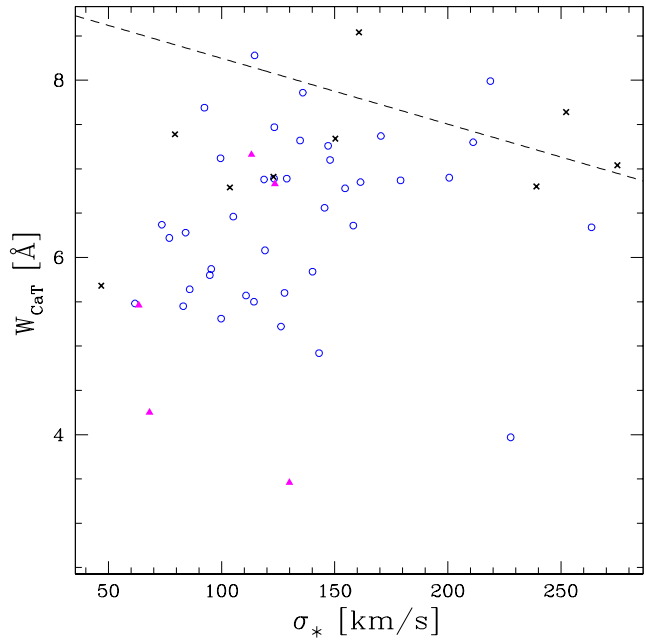}
\caption{CaT values versus stellar velocity dispersions. Dashed line
represents Saglia's fit (see text). Symbols as in
Fig. \ref{Fig_CaT_x_PL}.}
\label{Fig_CaT_x_Vd_all}
\end{figure}

Fig. \ref{Fig_CaT_x_Vd_all} shows the results of this test. Clearly,
as a whole, our galaxies do not follow a well defined $W_{\rm
  CaT}$-$\sigma_{\star}$ relation at all. The few normal galaxies in
our sample (crosses), as well as the high-$\sigma_{\star}$ Seyfert 2s,
do tend to follow a similar anti-correlation found by Saglia et al
(2002) for ellipticals. However, there is a large population of
galaxies with low-$\sigma_{\star}$ and relatively low $W_{\rm CaT}$
which has no correspondance in previous studies of this
relation. Excluding Mrk 3 and NGC 7212 (which are diluted by FC, as
discussed in \ref{Mixed Models}), the most deviant points are the
Starbursts NGC 1140, NGC 3256 and NGC 7714, located in the bottom left
part of the plot. Next, at $W_{\rm CaT}$ between 5 and 6.5 \AA\ and
$\sigma_{\star} <$ 150 km/s, are the composite Starburst-Seyfert 2s
nuclei, like NGC 5135, NGC 7582, IC 3639, Mark 1. It thus seems that
the complex star formation histories of these objects are somehow
related to their displacement with respect to the $W_{\rm
  CaT}$-$\sigma_{\star}$ relation defined for more well behaved
systems. Interestingly, the normal galaxy NGC 205 is located away the
anticorrelation stated for normal nuclei. This dwarf Elliptical galaxy
has a low metallicity (Mateo, 1998), which indicates that it could
also influence the CaT-$\sigma_{\star}$ relation.

\subsection{Stellar vs. NLR Kinematics}

\label{subsec:Stellar vs. NLR Kinematics}

The question concerning the acceleration of gas in the NLR has been
analyzed in several papers (NW, TDT). As pointed out by Green \& Ho
(2005; GH), the NLR is small enough to be illuminated by the active
nucleus and large enough to feel the gravitational forces of the bulge
of the host galaxy. Comparison between gas and stellar kinematics was
previously carried out by NW using the [OIII]$\lambda$5007 line, who
find that $\sigma_\star$ and the FWHM of [OIII]$\lambda$5007 are
correlated, but with substantial scatter. Recent work on Seyfert 1s
(Jim\'enez-Benito \etal 2000) and SDSS galaxies (Heckman \etal 2004,
GH) confirms this finding. The implication is that the NLR clouds are
at least partly dominated by virial motions in the host galaxy's
bulge. Given the existence of this relation, forbidden line widths
have sometimes been used as a surrogate for $\sigma_\star$ in the
absence of information on the stellar dynamics (Nelson 2000; Grupe \&
Mathur 2004). The case of Narrow-Line Seyfert 1 galaxies (NLS1) is
particularly interesting. In a recent paper, Komossa and Xu (2007)
have shown that, for these Seyfert galaxies, $\sigma_{\rm [OIII]}$ can
be used as a proxy for $\sigma_{\star}$ only after removing the
(usually blue) wings of the [OIII] profile. In other words,
$\sigma_{\rm [OIII]}$ should correspond only to the core of the line,
similarly to what we have done in the present paper in the fit of the
[SIII] emission-line profiles to obtain $\sigma_{\rm [SIII]}$.

\begin{figure}
\includegraphics[width=9cm]{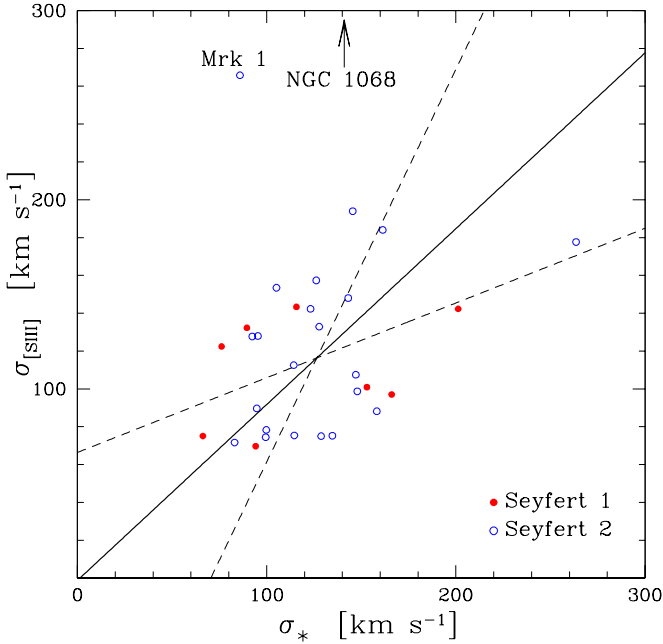}
\caption{Comparison of stellar velocity dispersions with our estimate
  of $\sigma_{\rm gas}$ based on the width of the [SIII]$\lambda$9069
  line. Symbols as in Fig. \ref{Fig_CaT_x_PL}. Ordinary Least Squares
  regressions are applied over all data but Mrk 1 and NGC 1068: solid,
  $OLS$ Bisector fit; dashed, $OLS(X|Y)$ and $OLS(Y|X)$ (shallowest
  and steepest lines, respectively).}
\label{Fig_Vdsiii_x_Vdstar}
\end{figure}

In Fig. \ref{Fig_Vdsiii_x_Vdstar} we compare our values of
$\sigma_{\rm [SIII]}$ and $\sigma_\star$ for all Seyfert nuclei. We
note that two of them, namely Mrk 1 and NGC 1068, have $\sigma_{\rm
  [SIII]}$ much larger than $\sigma_{\star}$. Evidence of jets were
found in these galaxies, which may explain the acelleration of the gas
via shocks produced in the interaction within jets and NLR. Excluding
these two nuclei, we have fitted all the other galaxies by linear
relations given by Ordinary Least Square ($OLS$) regressions (Isobe et
al, 1990). Solid line represents the $OLS$ Bisector fit, while dashed
lines are $OLS(X|Y)$ and $OLS(Y|X)$ (shallowest and steepest lines,
respectively). The OLS Bisector fit gives a slope of 0.93 (N = 29,
$R_{S}$ = 0.37, P = 0.046). Taking into account just Seyfert 2s, we
find a slope of 1.03 (N = 21, $R_{S}$ $\sim$ 0.40, P = 0.07) and 0.73
for Seyfert 1s, for which there is not a clear correlation ($R_{S}$
$\sim$ 0.31, P = 0.45). The slopes and correlation coefficients do not
vary substantially when we exclude worst quality data. Using the $a =$
very good, $b = $ good, and $c =$ not-too-bad quality flags adopted in
Paper I, we find for $a+b$ a slope of 1.00 (N = 23, $R_{S}$ = 0.39, P
= 0.06), and 1.00 for $a$ sources alone (N = 20, $R_{S}$ = 0.3, P =
0.2). All these is summarized in Table
\ref{OLS_for_[SIII]}. Interestingly, there is also a tendency for 4
Starburst nuclei, with a slope of about 0.56 and $R_{S}$ $\sim$ 0.81
(not shown here).

\begin{table}
\begin{center}
\begin{tabular}{lccccccl}
\multicolumn{6}{c}{} \\
\hline
                 & All   & Sy 1  &  Sy 2 &   All ($a+b$) &   All ($a$) \\ \hline
Slope            & 0.93  & 0.73  &  1.03 &   1.00     &   1.00    \\
N                & 29    &  8    &   21  &    23      &    20     \\
$R_{S}$          & 0.37  & 0.31  &  0.40 &   0.39     &   0.30    \\
P                & 0.046 & 0.45  &  0.07 &   0.06     &   0.20    \\
\end{tabular}
\end{center}
\caption{Results of OLS Bisector fits for
  $\sigma_{\rm [SIII]}$-$\sigma_{\star}$. Letters in columns 5 and 6
  denotes quality flags (See text).}
\label{OLS_for_[SIII]}
\end{table}

We repeated this analysis using $\sigma_{\rm_{[OIII]}} = {\rm
  FWHM}_{\rm [OIII]} / (8 \ln 2)^{1/2}$ taken from the literature
(Whittle, 1992; Bassani \etal 1999; Schmitt \etal 2003). NGC 1068, Mrk
1 and Mrk 78 present very different values of $\sigma_{\rm_{[OIII]}}$
and $\sigma_{\star}$. Further excluding the worst quality data (``flag
$c$'' in Whittle 1992 and our flag $d$) the OLS Bisector fit gives a
slope of 1.48 ($R_{S} = 0.42$) for the whole sample (28 objects),
while splitting into Seyfert 1s and 2s the slopes are 1.13 ($R_{S} =
0.81$) and 1.60 ($R_{S} = 0.33$) respectively. This is described in
Table \ref{OLS_for_[OIII]}. We point out that for this analysis we
used FWHM data of [OIII] line, which are (in median) larger than our
estimates of $\sigma_{\rm gas}$ through the FWHM of the $core$ of
[SIII] line, so that the slope of a fit applied to [OIII] data yields
larger values than those obtained for [SIII] data. We confirm this
fitting $\sigma_{\rm_{[SIII]}}$ vs $\sigma_{\rm_{[OIII]}}$, which
gives a slope of $\sim$ 0.85 ($R_{S}$ $\sim$ 0.44, 20 objects).

\begin{table}
\begin{center}
\begin{tabular}{lccccccl}
\multicolumn{4}{c}{} \\
\hline
                 & All  &  Sy 1 &   Sy 2  \\ \hline
Slope            & 1.48 &  1.13 &   1.60  \\
N                & 28   &  12   &   16    \\
$R_{S}$          & 0.42 &  0.81 &   0.33  \\
\end{tabular}
\end{center}
\caption{Results of OLS Bisector fits for $\sigma_{\rm
    [OIII]}$-$\sigma_{\star}$}.
\label{OLS_for_[OIII]}
\end{table}

\begin{figure}
\includegraphics[width=9cm]{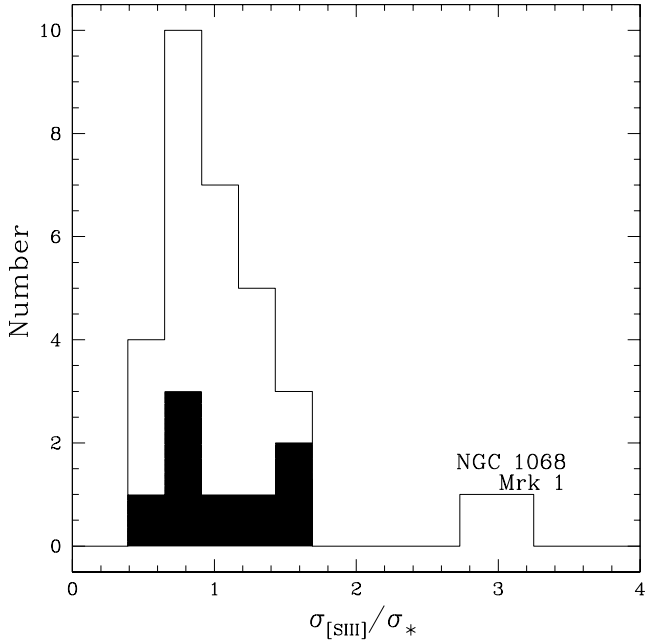}
\caption{Distribution of gas to stellar velocity dispersions ratio,
$\sigma_{\rm_{[SIII]}}$/$\sigma_{\star}$. Areas as explained in
Fig. \ref{Fig_hist_EWs}.}
\label{Fig_histogram_VDratio}
\end{figure}

There are, therefore, correlations between NLR and stellar motions,
but with considerable scatter. One way to study this scatter is to
look at the residuals $\sigma_{\rm_{[SIII]}}$/$\sigma_{\star}$ about
unity. Fig. \ref{Fig_histogram_VDratio} shows this distribution, which
is a moderately peaked (kurtosis $\sim$ -1) gaussian for the majority
of Seyfert 2s. Without taking into account the two galaxies with
abnormally high gaseous velocity dispersion (for which it was shown
that there are several kinematical components),
$\sigma_{\rm_{[SIII]}}$/$\sigma_{\star}$ = 0.97 $\pm$ 0.32 (average
and rms dispersion), while for Seyfert 1s and 2s
$\sigma_{\rm_{[SIII]}}$/$\sigma_{\star}$ are 1.02 $\pm$ 0.40 and 0.95
$\pm$ 0.29, respectively, i.e.,
$\sigma_{\rm_{[SIII]}}$/$\sigma_{\star}$ is the same for both types of
objects. Fig. \ref{Fig_histogram_VDratio} is similar to fig. 7 of NW,
which uses [OIII]-based gas kinematics. Although NW have twice as many
objects, we note that our distribution is more concentrated around
unity, mainly because we have FWHM determinations of gaussians fitted
to the {\em core} of the [SIII] line.

All this study shows that the [SIII] core emitting zone of the NLR is
dynamically linked to the gravitational bulge potential, a result
similar to that obtained by Komossa and Xu (2007) for NLS1
galaxies. However, we see that there are possible secondary factors
besides gravitational potential that could explain the scatter about
unity in NLR kinematics. Shocks and other acceleration mechanisms
could account for the high velocity clouds observed in a few galaxies,
but in general a complete study of gas-stellar kinematics remains to
be done.

\begin{figure}
\includegraphics[width=9cm]{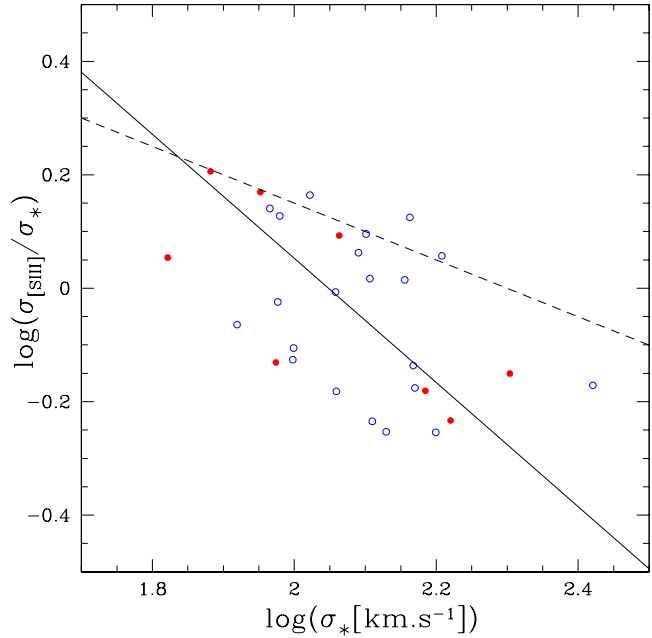}
\caption{Deviation from $\sigma_{\rm_{[SIII]}}$/$\sigma_{\star}$ with
  $\sigma_{\star}$. Solid line is the least square fit for all
  Seyferts. Dotted line, Green\&Ho's fit. Symbols as in
  Fig. \ref{Fig_CaT_x_PL}.}
\label{Fig_RatioVd_x_Vdstar}
\end{figure}

A detailed analysis of stellar and gaseous kinematics of NLR was made
by GH taking precisely $\sigma_{\rm_{[SIII]}}$/$\sigma_{\star}$ as the
central quantity of their study. They found that none but one of the
nuclear activity indicators is correlated with
$\sigma_{\rm_{[SIII]}}$/$\sigma_{\star}$; only with $\log
\sigma_{\star}$, which is indirectly a measure of bulge mass, there is
a hint of relationship. Although we have few points, we note this
trend in our data, which is shown in Fig.\ \ref{Fig_RatioVd_x_Vdstar}
together with GH's best fit. Again, applying OLS Bisector fit, we
obtain $\log \sigma_{\rm_{[SIII]}}$/$\sigma_{\star} = -1.1 \log
\sigma_{\star} +2.2$, with $R_{S} = -0.55$. A least squares fit gives
$\log \sigma_{\rm_{[SIII]}}/\sigma_{\star} =-0.56 \log \sigma_{\star}
+ 1.13$ ($R_{S}$ = -0.55, P = 0.022), nearly the same relation
obtained by them. Only for Seyfert 2s there is a weaker (but real)
trend, $\log \sigma_{\rm_{[SIII]}}/\sigma_{\star}=-0.41 \log
\sigma_{\star}+0.81$ (N = 21, $R_S = -0.37$, P = 0.09), while for
Seyfert 1s the correlation is moderately strong, $\log
\sigma_{\rm_{[SIII]}}/\sigma_{\star}=-0.74 \log\sigma_{\star}+1.50$ (N
= 8, $R_S = 0.75$, P = 0.036).

We checked whether $\sigma_{\rm_{[SIII]}}/\sigma_\star$ correlates
with other properties, like morphological and activity types,
redshift, inclination, etc., but (like GH) found no strong trend. The
absence of a trend with the inclination of the host galaxy suggests
that the gas in the NLR has random velocities and so the velocity
field is similar to that of the stars, as previously found by NW
(their Fig.\ 11).

Our data thus support the existence of an anticorrelation between
$\sigma_{\rm [SIII]}$ and $\sigma_{\star}$, albeit with a large
scatter.  Using $sigma_{\rm_{[SIII]}}$ as a proxy of $\sigma_{\star}$
tends to underestimate velocity dispersion in bulge of massive
galaxies. In any case, given the existence of a correlation between
both (gas and stellar) velocity dispersions, and taking into account
that [SIII] is a high ionization line produced very near the active
nucleus and often with strong wings like [OIII], our study supports
the idea of using $\sigma_{\rm_{[SIII]}}$ as a kinematic tracer of the
NLR.

\subsection{Spatially resolved kinematics}

As an extension of our spatial analysis, we report results for the
variations in kinematical properties of our sub-sample.

\begin{figure}
\includegraphics[width=0.5\textwidth, bb=18 144 420 575]{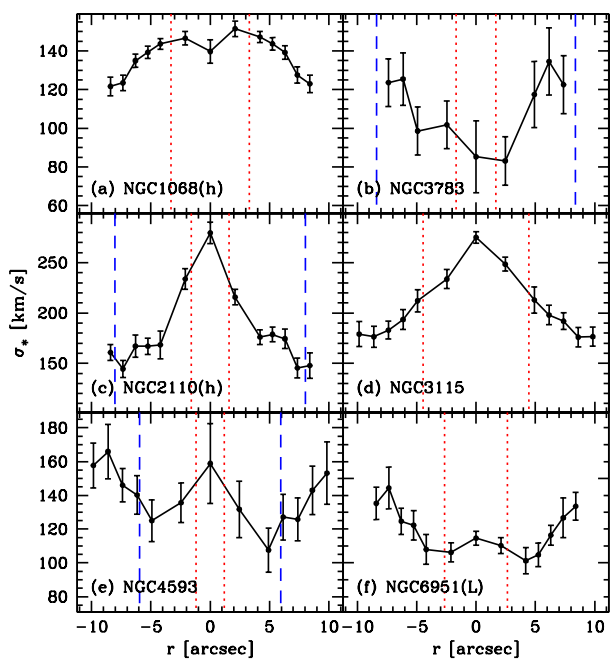}
\caption{Examples of $\sigma_\star$ spatial variations. Dotted lines
  mark $-200$ to +200 pc region; dashed lines, the $-1$ to +1 kpc
  region.}
\label{fig:Spatial_Sig}
\end{figure}

Fig. \ref{fig:Spatial_Sig} shows examples of spatial variations in
$\sigma_\star$. Some recent works (Barbosa \etal (2006),
Garc\'ia-Lorenzo \etal 1999; Emsellen \etal 2001; M\'arquez \etal
2003) have discussed the possibility of a {\em sigma-drop} detection,
i.e., that $\sigma_{\star}$ is appreciably lower towards nuclear
regions due to the presence of a star-forming cold nuclear disc. We
see a nuclear drop in $\sigma_{\star}$ in only two galaxies
(Fig. \ref{fig:Spatial_Sig}a and b). For some objects there is even a
hint of a rise in $\sigma_{\star}$ in their central parts (e.g.,
Fig. \ref{fig:Spatial_Sig}c and d). We have to keep in mind, however,
that our spectra do not have the spatial resolution needed to detect
this behavior in $\sigma_{\star}$, which would require a resolution of
a few tens parsecs.

Another interesting feature that can be seen for some of our objects
is a symmetric off-nuclear drop in $\sigma_\star$, as in NGC 4593
(Fig. \ref{fig:Spatial_Sig}e). This is in agreement with the detailed
integral field spectroscopy of this object made by Barbosa \etal
(2006), who found a partial ring of low-$\sigma_{\star}$ nearly at the
same distance from the nucleus. A similar behavior is seen in NGC 6951
(Fig. \ref{fig:Spatial_Sig}f). U-band images of these galaxies
(Mu\~noz Mart\'{\i}n et al, 2007) shows the presence of young stars at
the locations where the off nuclear $\sigma_{\star}$ drops are
observed.

\section{Summary}

\label{sec:Summary}

In Paper I we presented a spectroscopic atlas of 78 galaxies in the
region of CaT lines, and measured the equivalent widths of CaT and
[SIII]$\lambda$9069 lines, and stellar and gaseous velocity
dispersions. In this Paper we used these data, as other measurements,
to investigate the CaT strength and the kinematical properties of
Seyfert nuclei.

Our conclusions can be summarized as follows:

\begin{enumerate}

\item The equivalent widths of CaT and [SIII] are diluted in Seyfert
  1s, due to the presence of a non-stellar component, while in Seyfert
  2s there is no sign of dilution. We show that the nuclear dilution
  of CaT lines in Seyfert 1s and the non observed dilution in Seyfert
  2s holds spatially, i.e., within approximately the central kpc. At
  this distance the effects disappears, thus implying that the stellar
  lines are diluted due the nuclear non-stellar continuum directly
  seen in Seyfert 1s.

\item The CaT strength turns out not to be a good tracer for the mean
  ages and metallicities of metal rich stellar populations.
  Nonetheless, its combination with CaK line yields a useful
  constraint on the nature of the continuum emission from optical to
  near-IR wavelengths.

\item We show that the location of Seyfert 2s in the CaT-CaK plane
  could be explained satisfactorily by stellar population synthesis
  models, by considering mixtures of an old plus very young stellar
  populations with a extinction of just two or three magnitudes. The
  hypothesis of Seyfert 2 nuclei composition as a mixing of an old
  stellar population plus power-law central source cannot account for
  the majority of the data.

\item There is a correlation between nuclear $\sigma_\star$ and
  $\sigma_{\rm [SIII]}$, as well as an anticorrelation between
  $\sigma_{\rm [SIII]}$/$\sigma_{\star}$ and $\sigma_{\star}$, both
  with substantial scatter. This means that care must be exercised
  when using $\sigma_{\rm [SIII]}$ as a proxy for
  $\sigma_\star$. These results are compatible with previous results
  of NW and GH.

\end{enumerate}

\section*{ACKNOWLEDGMENTS}

LRV, NVA, RCF, AGR and TSB acknowledge the support of Capes and
CNPq. LRV acknowledge the support from Secyt, and the hospitality of
UFSC. RGD acknowledges support by Spanish Ministry of Science and
Technology (MCYT) through grant AYA-2001-3939-C03-01. We thank
Laborat\'orio Nacional de Astrof\'isica for the allocation of time on
ESO 1.52m and financial support during the runs. We also thank the
referee, Javier Cenarro, for the very useful comments which greatly
improved the content of the paper. Part of the data described here
were taken at Kitt Peak National Observatory, National Optical
Astronomy Observatories, which are operated by AURA, Inc., under a
cooperative agreement with the National Science Foundation. Basic
research at the NRL is supported by 6.1 base funding. This research
made use of the NASA/IPAC Extragalactic Database (NED), which is
operated by the Jet Propulsion Laboratory, Caltech, under contract
with NASA.

\end{document}